\documentclass[a4paper,fleqn,usenatbib,twocolumn]{mnras}
\usepackage{newtxtext,newtxmath}
\usepackage[T1]{fontenc}
\usepackage{ae,aecompl}
\usepackage{graphicx}
\usepackage{amsmath}	
\usepackage{amssymb}	
\usepackage{subfig}
\usepackage{multirow}
\usepackage{tabularx}
\usepackage[dvipsnames]{xcolor}
\usepackage{array}
\usepackage{booktabs} 
\newcolumntype{P}[1]{>{\centering\arraybackslash}p{#1}}
\newcommand{\be}{\begin{equation}}
\newcommand{\ee}{\end{equation}}
\usepackage{calligra} 
\DeclareMathAlphabet{\mathcalligra}{T1}{calligra}{m}{n} 
\DeclareFontShape{T1}{calligra}{m}{n}{<->s*[2.2]callig15}{}

\newcommand{\mach}{\mathcal{M}}

\title[IMRIs in gas discs]{Probing gas disc physics with LISA: simulations of an intermediate mass ratio inspiral in an accretion disc}

\author[A. Derdzinski]{A.~M.~Derdzinski$^{1}\thanks{E-mail: aderdzinski@astro.columbia.edu}$, D. D'Orazio$^{2}$, P. Duffell$^{2}$,  Z. Haiman$^{1}$, A. MacFadyen$^{3}$\\
$^{1}$Department of Astronomy, Columbia University, New York, NY, 10027, USA\\
$^{2}$Department of Astronomy, Harvard University, 60 Garden Street Cambridge, MA 01238, USA\\
$^{3}$Center for Cosmology and Particle Physics, Physics Department, New York University, New York, NY 10003,USA}

\begin{document}
\date{Received / Accepted}
\pagerange{\pageref{firstpage}--\pageref{lastpage}} \pubyear{2017}

\maketitle
\label{firstpage}

\begin{abstract}
  The coalescence of a compact object with a $10^{4}-10^{7} {\rm
    M_\odot}$ supermassive black hole (SMBH) produces mHz
  gravitational waves (GWs) detectable by the future Laser
  Interferometer Space Antenna (LISA). If such an inspiral occurs in
  the accretion disc of an active galactic nucleus (AGN), the gas torques
  imprint a small deviation in the GW waveform.
  Here we present two-dimensional hydrodynamical simulations with the
  moving-mesh code DISCO of a BH inspiraling at the GW rate in a
  binary system with a mass ratio $q\!=\!M_2/M_1\!=\!10^{-3}$,
  embedded in an accretion disc.  We assume a locally isothermal
  equation of state for the gas (with Mach number $\mach=20$) and
  implement a standard $\alpha$-prescription for its viscosity (with
  $\alpha = 0.03$).
  We find disc torques on the binary that are weaker than in
  previous semi-analytic toy models, and are in the opposite
  direction: the gas disc slows down, rather than speeds up the
  inspiral.  We compute the resulting deviations in the GW waveform,
  which scale linearly with the mass of the disc.
  The SNR of these deviations accumulates mostly at high frequencies,
  and becomes detectable in a 5-year LISA observation if the total
  phase shift exceeds a few radians.  We find that this occurs if the
  disc surface density exceeds $\Sigma_0 \gtrsim 10^{2-3}\rm
  g\,cm^{-2}$, as may be the case in thin discs with near-Eddington
  accretion rates.
  Since the characteristic imprint on the GW signal is strongly
  dependent on disc parameters, a LISA detection of an intermediate mass ratio inspiral would probe the
  physics of AGN discs and migration.

\end{abstract} 
  
\begin{keywords}
accretion / accretion discs, black hole physics, gravitational waves, hydrodynamics
\end{keywords}
\section{Introduction} 
\label{sec:introduction}

LISA is currently planned to launch by 2034 and is expected to herald the era of
space-based interferometry with the detection of gravitational waves
(GWs) at wavelengths larger than the Earth.  
With an interferometer
arm-length of 2.5 million km, LISA will probe the mHz GW sky with the
primary goal of detecting merging supermassive black holes
(SMBHs) throughout cosmic history.  While the loudest sources in the
LISA frequency band include merging SMBHs with component masses
$M_{\rm BH}\!\sim\! 10^4\!-\!10^7 M_{\odot}$ up to a redshift
$z\!\sim\!20$, LISA will also be sensitive to less massive compact
objects coalescing with SMBHs. These events are referred to as
intermediate mass ratio inspirals (IMRIs; $q \equiv
M_2/M_1\!\approx\!10^{-3}\!-\!10^{-4}$) or extreme mass ratio
inspirals (EMRIs; $q\!\lesssim\!10^{-4}$, detectable up to
$z\!\sim\!4$, \citealt{Seoane2017}).

Unlike stellar-mass BH mergers which are presumed to occur in vacuum
(although see \citealt{Perna+2016, Janiuk+2017, Bartos2017,McKernan2017,Stone2017,
D'OrazioLoeb:2018}), many LISA
events may occur in gaseous environments in galactic nuclei. Only
$\sim\!1\%$ of galaxies host active galactic nuclei (AGN), in which
the central SMBH is accreting at nearly the Eddington rate from a
thin, cold accretion disc, but there is plenty of evidence that AGN
are triggered by galactic mergers (e.g. \citealt{Kauffmann2000,
  Hopkins2008, Goulding2018} and references therein). In particular,
the merger of two massive galaxies results in a supply of gas that
flows into the nucleus of the post-merger remnant
\citep{DottiSesanaDecarli2012,BarnesHernquist1996},
providing a gas-rich environment for BH accretion. 
As a result, a large fraction of SMBH mergers are expected to occur in a
gaseous environment (see, e.g. \citealt{mayer2013} for a review).

If a coalescing compact BH binary encounters a sufficient amount of
gas, it will experience a gravitational torque that can act to either
accelerate or hinder a GW-driven inspiral. The presence of gas also
provides the opportunity for BHs to accrete, which in turn will affect
their mass, spin, and momentum.  This raises an important question:
could gas change the orbital evolution of a binary sufficiently
strongly such that the corresponding changes in the GW waveform become
measurable? If it does, the GW signal would not only provide
information about the source parameters, but it also would carry a
characteristic signature of the environment in which the source
originated.

Thus far the impact of gaseous forces on GW signals has been addressed
only via semi-analytical toy models \citep{Kocsis2011, Yunes2011,
  Barausse+2014, BarausseCardosoPani2015}.  The overall conclusion
from these studies is that gas has a negligible impact for SMBH
binaries in the LISA band, except for systems with extreme mass ratios
($q\!\ll\!1$). Environmental influences become important for sources
with less massive companions, where GWs are weaker and gas effects are
comparatively stronger.  Semi-analytical estimates by
\citealt{Kocsis2011} focused on EMRIs and found that a dense,
near-Eddington gas disc can speed up the inspiral to an extent
observable by LISA.
     
GW sources with $q\!\ll\!1$ include a range of possible SMBH component
masses, but in the present study we focus on IMRIs that will fall in
the mHz GW band. The two relevant cases in this regime include the
mergers of massive stellar remnant BHs ($M_{\rm BH}\!\sim\!10-100
M_{\odot}$) with IMBHs ($M_{\rm BH}\!\sim\!10^{4-5} M_{\odot}$), or
$10^{3-4}M_{\odot}$ IMBHs coalescing into $10^{6-7}M_{\odot}$
SMBHs. For our detectability estimates we adopt the specific case of
a $q = 10^{-3}$ mass-ratio binary with a primary BH mass of $M_1=10^6
M_{\odot}$.

IMRI rate estimates predict a few to tens of mergers per year in the universe
\citep{AmaroSeoane2007, Miller2009}.
These estimates are based on stellar dynamical processes in galactic nuclei,
and only a small fraction of these events would be expected to occur in a gaseous
environment. However, IMRIs could also occur in AGN discs via several additional
evolutionary pathways,
either from compact objects in the galactic nucleus whose orbits are
dragged into the plane of the disc by repeatedly crossing the disc
(e.g. \citealt{ipp99}; \citealt{McKernan2012}; \citealt{Kennedy2016} and references therein), or
from compact remnants that are formed in the disc in the first place
(e.g. \citealt{GoodTan2004, Levin2007, McKernan2014}; see also
\citealt{Stone2017}).  Subsequent accretion and mergers
of these remnants can lead to IMBHS embedded in AGN discs
(e.g. \citealt{Bellovary2016,yi+2018}).
Similarly to near equal-mass SMBH mergers, 
E/IMRIs may preferentially occur in AGN discs.

The previous studies mentioned above estimate the gas impact on
E/IMRIs in near-Eddington accretion discs with semi-analytical models
of the so-called migration torque.
For IMRIs in particular, these migration torques are
based on the viscous torque and
are estimated for a non-inspiraling perturber on a fixed circular orbit.
However, a rapidly inspiraling, GW-driven perturber modifies the disc
structure differently from a non-migrating perturber. As a result, the
torques for a moving perturber differ in both strength and sign
from the torques for a perturber on a fixed orbit \citep{Duffell2014}, and should also evolve differently during the inspiral.

Motivated by the above, in this paper we perform high--resolution
two--dimensional (2D) hydrodynamical simulations, and we directly measure
the torques exerted on an IMBH embedded in the accretion disc
of a central SMBH.
We use the moving-mesh grid code DISCO \citep{Duffell2016}, model the
IMBH as a sink particle, and assume its orbit follows a GW-driven inspiral.
The disc is assumed to have a locally
isothermal equation of state and to obey a standard
$\alpha$-prescription for its viscosity.
We calculate the impact of the disc torques throughout the
coalescence, and predict the corresponding modification to the GW
waveform seen by LISA. We also compute the signal-to-noise ratio (SNR)
of the detectability of these modifications.

Focusing on the
intermediate mass-ratio regime has two advantages 
compared to EMRIs.  First, the inspiral is more rapid, allowing us to
simulate a large portion of the inspiral as it traces out a broad
range of orbital separations and frequencies.  Second, the system is
easier to resolve numerically, allowing us to follow the system for as
many as $\approx 10,000$ binary orbits.

Our primary goal is to estimate the detectability of the gas imprint
on an IMRI with the currently proposed LISA configuration
\citep{Klein2016}.  The torques scale linearly with the mass of the
AGN disc involved, and also depend on other disc properties, such as
temperature and viscosity. Therefore, a measurement of a gas imprint
on the inspiral waveform should probe the properties of the accretion
disc in which the source resides.

This paper is organised as follows.
In \S~\ref{sec:previous}, we summarise previous work on migration
torques in more detail.
In \S~\ref{sec:methods}, we describe our simulation setup, and 
in \S~\ref{sec:results} we present our results, focusing on the torque
measurements.
In \S~\ref{sec:detectability} we use the measured torques to compute
the modifications of the GW signal and estimate their detectability
with LISA.
In \S~\ref{sec:discussion}, we discuss our results, along with some
caveats, and finally
in \S~\ref{sec:summary} we summarise our conclusions and the
implications of this work.

\section{Previous work on migration torques for GW sources}
\label{sec:previous}

Early estimates of the impact of a gas disc on the gravitational
waveforms of a compact object (CO) spiraling into a SMBH appear in
\citet{Chakrabarti1996} and \citet{Narayan2000}.
These studies focused
on the angular momentum exchange between the CO and the disc due to
accretion and hydrodynamical drag in disc models at different
radiative efficiencies. \citet{Levin2007} considered, additionally, the
impact of torques from density waves excited in a thin accretion
disc~\citep{GT80}, often called ``Type I torques'' in the context of
extrasolar planets (see below).  The impact of a broader range of
environmental effects on EMRI gravitational waveforms has been
enumerated in \citet{Yunes2011} and \citet{Barausse+2015}.

These studies found that the effect of gas is generally very weak and
undetectable for LISA sources.\footnote{It is worth noting that gas
  discs could be much more important, and produce order-unity effects,
  for the more massive, sub-parsec separation $\sim\!10^9~{\rm M_\odot}$
  SMBH binaries at nano-Hz frequencies, detectable by pulsar timing
  arrays \citep{KocsisSesana2011,TMH12,Sesana+12}.}  The exceptions
are systems with a small ratio, for which the gas torques due to the
tidal deformation of the disc (so-called migration torques) can
produce detectable deviations to the GW torque.  Such disc torques
are more extensively studied in the context of
protoplanetary discs and planet migration, including numerous two- and
three-dimensional hydrodynamical simulations (see a review by,
e.g. \citealt{Baru2014}).

In the protoplanetary disc context, migration is described by two limiting
cases (Type~I and Type~II; \citealt{ward97}). Type~I torques on low
mass-ratio planet systems ($q\!<\!10^{-4}$) are well understood, in the sense that
they can be described by a linear perturbation theory~\citep{GT80},
which is successfully reproduced in numerical studies
(e.g. \citealt{Tanaka2002}). However, these torques can be sensitive
to disc thermodynamics (\citealt{Paard2006}).  Type~II migration
concerns more massive planets that can carve gaps in their discs (for
intermediate masses $q\!>\!10^{-3}$, with the precise value depending
on the disc viscosity and temperature). In this regime the torques
become nonlinear, so there is no analytical solution for the migration
rate.  Often, semi-analytical estimates are based on the viscous
torque, but hydrodynamical simulations show that migration in this
regime can deviate significantly from the viscous rate
(\citealt{Edgar2007, Crida2007, Duffell2014, RobertCrida2018}). For
intermediate mass perturbers in particular, the torque develops a
nonlinear component that is remarkably sensitive to disc parameters
such as the viscosity, Mach number, and density gradient
\citep{Duffell2015}.

In the context of binaries embedded in AGN discs, \citet{Yunes2011}
and \citet{Kocsis2011} computed modified waveforms, using
semi-analytical formulae of the above planetary migration torques, for
both extreme and intermediate-mass perturbers.  \citet{Kocsis2011}, in
particular, showed that the gas-induced drift
in the accumulated GW phase can exceed a few radians per year
(depending on disc parameters), a limit that is detectable with LISA's
sensitivity. For the most massive disc models, the drift can become as
large as 1,000 radians per year, which provides the possibility for
LISA to accurately probe disc parameters (possibly limited by
degeneracies between migration and system parameters).

In our present study, we focus on the analog of gap-opening satellites
(for $q=10^{-3}$) in the context of BHs.  While similar to the planet
case, there are three major differences: (1) the BHs accrete, which is
typically ignored when calculating planetary torques, (2) the BH orbit
is strongly dominated by GWs (in the LISA band), there is no analogous
``external'' force in the planet case, and (3) the physical parameters
of AGN discs differ from protoplanetary discs, which are typically
thicker (lower Mach number) and have a lower viscosity.

The combination of these effects on migration torques has not been
studied in detail in any simulation to date.  For completeness, we
note that related works exist, which present simulations of gas discs
that include unequal-mass, GW-driven binaries, both in 1D
\citep{Chang2010,fontecilla+17,tl15} and in 3D
\citep{Baruteau2012,Cerioli+2016}.  These studies aimed at
understanding the tidal squeezing of the inner disc by the inspiraling
companion (a `snow-plow' effect), and the resulting enhanced accretion
rate onto the primary BH and electromagnetic (EM) emission.  These
works did not measure the gas torques on the binary.  Likewise, recent
2D simulations followed the GW-driven inspiral of an equal-mass SMBH
binary~\citep{Farris2015,Tang2018} in the LISA band, but focused on
the accretion rates and electromagnetic (EM) emission, and did not measure the torques
in this regime.

Finally, we note the work of \citet{Duffell2014}, whose results
directly inspired the present study.  These authors measured the
torques on a Jupiter-like planet ($q=10^{-3}$), embedded in a
protoplanetary disc, and migrating at a broad range of {\em manually
  prescribed} rates.  By measuring the torques as a function of the
rate at which the planet is dragged inward, a unique migration rate
can be identified that is consistent with the torques measured at that
rate.  This is meant as a ``trick'' to avoid a numerically
challenging simulation of a ``live'' binary coupled to the disc.
However, as a by-product, \citet{Duffell2014} inadvertently studied
migration in a case similar to GW-driven inspiral.  They found that
above some migration rate [$a(da/dt)^{-1}\gtrsim 10^{5} t_{\rm
  orb}\sim t_{\rm visc}$, where $a$ is the binary separation, $t_{\rm
  orb}$ is its orbital period, and $t_{\rm visc}$ the viscous
timescale], the gas torques are modified.  Interestingly, at
sufficiently rapid migration rates, the torques change sign and start
to slow down, rather than speed up, the inspiral. While their study
 focused on non-accreting planets in protoplanetary discs, the
dependence of torques on migration rate implies that the gas torques
of a non-migrating secondary cannot simply be linearly added to the
much larger external torques (in our case, from GWs).

In summary, in the present paper, we extend semi-analytic estimates of
the modified GW waveforms \citep{Kocsis2011}, by running hydrodynamic
simulations with a setup similar to \citet{Duffell2014}, except with
the addition of accretion, GW-driven migration, and a more AGN-like
disc model.  We expect that the torques we measure will be very
sensitive to disc parameters, and therefore this study is but the
first step toward a complete exploration of the importance of gas
discs for LISA sources (Derdzinski et al, in prep).

\section{Numerical methods} 
\label{sec:methods}

\subsection{Disc Model}

Our model is a two-dimensional, locally isothermal, viscous gas disc,
evolved using the moving-mesh grid code DISCO
\citep{Duffell2016}. DISCO is idealised for modeling accretion discs
as it has the capability to allow the grid to move with the Keplerian
flow of the gas, thus reducing advection errors produced by shearing
flow between grid cells.

The computation domain extends from $0.5\!\le\!r/r_0\!\le\!2.75$ where
 $r$ is measured from the primary BH which is held at the origin, and $r_0$ is an arbitrary distance unit set to $r_0=1$.
Likewise, the primary mass is set to $GM_1=1$ in code units (where $G$
is the gravitational constant).  Note that the orbital time in code
units at $r=r_0$ (the final binary separation) is $2\pi$.
The grid is logarithmically spaced in polar coordinates, with a
total of 512 radial cells and an increasing number of azimuthal cells
at outer radii such that the aspect ratio of each cell is unity.  This
resolution reaches 30 zones per scale height, which is important for
capturing the gas morphology around the secondary BH.  To ensure that
our resolution is adequate, we have performed a test run with 800
radial grid cells, and found that our measured torques do not change.
We employ Dirichlet (fixed) conditions at the inner and outer
boundaries to ensure a constant mass flux rate.

The disc is parameterised by a constant aspect ratio
$h/r\!=\!\mach^{-1}$, where $h$ is the disc scale height, $r$ is the
distance from the SMBH, and $\mach$ is the Mach number, assumed to be
independent of $r$.  Under this
assumption, gas dynamics is scale-free as the black hole migrates
through the domain, and the sound speed varies as $c_s = v_{\phi} /
\mach $, where $v_{\phi} = r \Omega$ is the orbital velocity and
$\Omega$ is the
initial Keplerian orbital frequency at $r$. 
Viscosity is set with an $\alpha$-law
prescription, with the kinematic viscosity provided by $\nu = \alpha c_s h$. We neglect radiative cooling and instead assume the disc
is locally isothermal by setting the vertically integrated pressure to $p = c_s^2 \Sigma(r)$, 
 such that the mach number and corresponding temperature profile remain fixed. 
Here $\Sigma(r)$ is the vertically-integrated
surface density, which is assumed to follow
\be
\Sigma(r) = \Sigma_0 \left(\frac{r}{r_0}\right)^{-1/2},
\ee
where $\Sigma_0\equiv\Sigma(r_0)$  is a constant scaling factor.
Since the orbit of the
secondary BH is imposed by hand (\S~\ref{sec:themigrator}) and we
have no self-gravity, $\Sigma_0$ can be scaled to any value.  Gas
forces on the BH simply scale linearly with $\Sigma_0$, as long as
$\Sigma_0$ remains low enough not to significantly modify the orbit.
This assumes that the evolution of the binary is overwhelmingly
dominated by GWs, which we show to be valid in \S~\ref{sec:GRtorque}
below.

For parameters describing the disc temperature and viscosity, we
choose $\mach\!=\!20$ and $\alpha\!=\!0.03$. While neither of these
values are typical of what we expect in AGN discs (where we believe
the gas is highly supersonic, with $\mach\!\sim\!100$ and highly
ionised leading to $\alpha\!\sim\!0.1\!-\!0.3$), we implement these
values here because they are (i) numerically easier to simulate and
(ii) this particular combination of $\mach$ and $\alpha$ leads to
a gap with a similar depth to a disc with AGN-like parameters.

\begin{figure}
\begin{center}
\includegraphics[width=.5\textwidth]{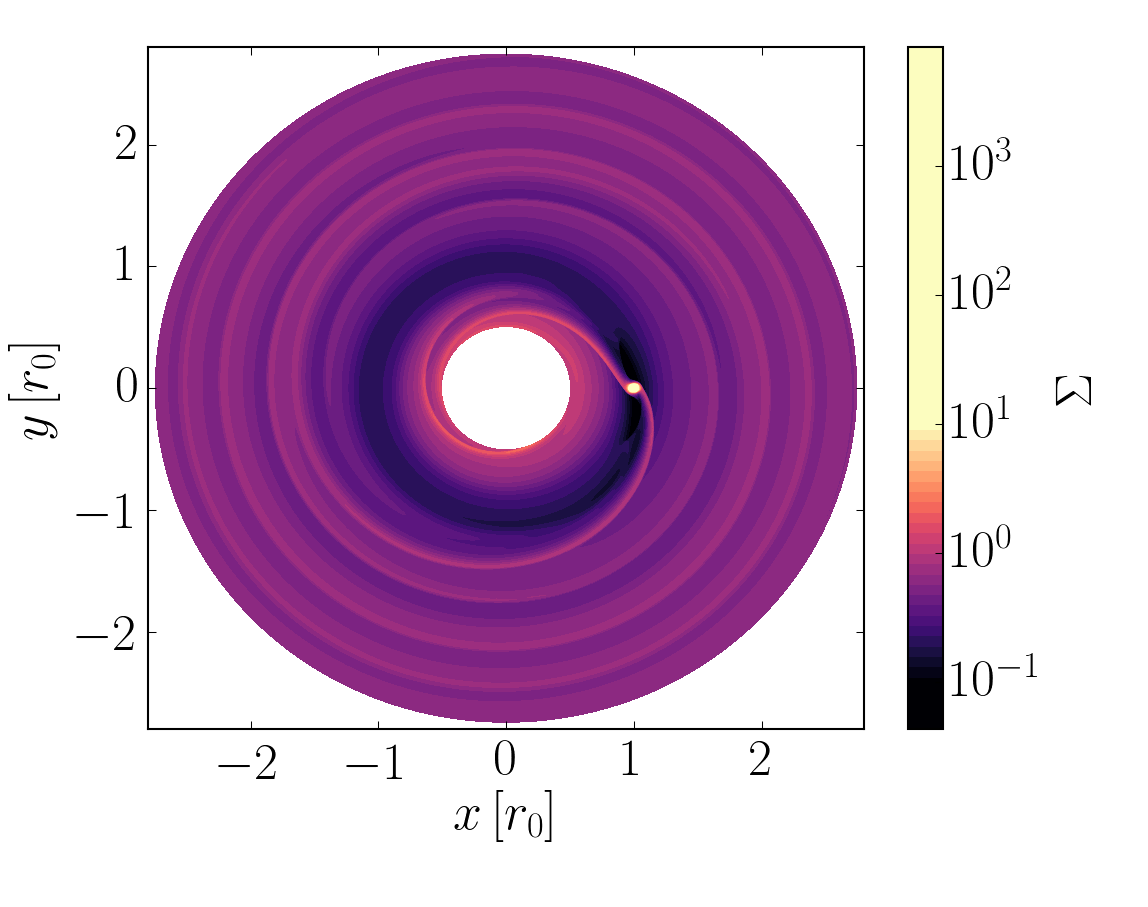} 
\caption{The logarithmic surface density over the whole computational domain,
  at the end of a simulation run.  
  The primary BH is at the origin, and the gas disc and the secondary BH are both orbiting counter-clockwise.
  The secondary BH is located at $(x,y)=(1,0)$, marked by a large overdensity.}
\label{fig:setup}
\end{center}
\end{figure}

Fig.~\ref{fig:setup} shows a snapshot of the logarithmic 2D surface
density over the whole computational domain, at the end of an
illustrative run.  

\subsection{The migrator and the GW inspiral}
\label{sec:themigrator}

The primary BH is held at the origin which is excised from the
simulation domain. The secondary BH is placed in a prograde orbit, and
modeled by a `vertically averaged' potential of the form
\be
\Phi_2 = \frac{G M_2}{(r_2\, {}^2+\epsilon^2)^{1/2}},
\ee
where $r_2$ is the distance to the secondary BH and $\epsilon$ is a
smoothing length, which we set to one half the scale height. 
 This approach essentially neglects the gravitational pull of the secondary from within 
  $r_2 < \epsilon < r/\mach$.
  The purpose of smoothing the potential is not only to avoid the
singularity at the position of the secondary, but to mimic the
vertically integrated forces that the two-dimensional fluid elements
feel within a scale height of the BH (see \citealt{Tanaka2002,
  Masset2002, Muller2012}).

While the disc is initially steady, it experiences perturbations in
response to the placement of the secondary BH. These perturbations are
transient and decay after a viscous time. The viscous time is
given by
\be
\label{eq:tvisc}
t_{\rm visc} (r) =  \frac{2}{3} \frac{r^2}{\nu} 
= \frac{\mach^2}{3 \pi \alpha} t_{\rm orb}(r)
\approx
1415 \,  \Bigg(\frac{\mach}{20}\Bigg)^2 \Bigg(\frac{\alpha}{0.03}\Bigg)^{-1} t_{\rm orb}
\ee
where $\nu = \alpha c_s h = \alpha h^2 \Omega,$ is the kinematic
viscosity, and we define $t_{\rm visc}$ as a function of orbital time
$t_{\rm orb}$ at the secondary location $r$.  We therefore disregard
the dynamics during the first $1400$ orbits of the simulation when
measuring the torques, to avoid including numerical transients. (Note
that the secondary only moves a distance of $\sim 0.1 r_0$ during this
phase.)

For a binary being driven together by gravitational waves, the
quadrupole approximation (\citealt{Peters64}) for the evolution of the
orbital separation is
\begin{equation}
\dot{r}_{\rm GW} = -\frac{64}{5}\frac{(GM)^3}{c^5}\frac{1}{1+q^{-1}}\frac{1}{1+q} \frac{1}{r^3},
\end{equation}
where $G$ is the gravitational constant, $c$ is the speed of light,
and $M=M_1+M_2$ is the total binary mass.  Integrating this
expression, the secondary's position can be written as
\be
\label{eq:GWpos}
r(t) = r_{\rm min} \left[1 - 4 R (t-t_{\rm total}) \right]^{1/4},
\ee
where $t$ is the elapsed time, $t_{\rm total}$ is the total simulation
time, $r_{\rm min}$ is the final separation at $t=t_{\rm total}$, and
$R \equiv \dot{r}_{\rm GW}/r$ is the inspiral rate defined at the
final separation.  In principle, we have the choice of specifying the
initial and final position of the secondary in code units, as well as
the physical scale for the total mass $M$ (note that $q=10^{-3}$ has
already been fixed). In practice, we are numerically limited by the
total number of orbits we can simulate ($\approx 10,000$ at our chosen
resolution). Our choice for, say $r_{\rm min}$ and the physical mass
scale, is further constrained in order for the binary to be chirping
(i.e. changing its separation noticeably during the simulation), and
for the GW frequency to fall in the LISA band.  We therefore chose
parameters that are appropriate for a LISA IMRI: $M_1 = 10^6 M_{\odot}$,
$q = 10^{-3}$, $r_{\rm min} = 5 r_{\rm S}$ (where $r_{\rm S} = 2GM_1/c^2$ is
the Schwarzschild radius). Our choice of
covering $\sim\!10,000$ orbits leads to an initial position of the
secondary BH being $r_{\rm max} = 11 r_S$.  Simulation parameters are
defined in Table~\ref{table:parameters}.  Note that in code units, 
$r_{\rm min}=1$ (i.e. twice the inner boundary), $r_{\rm max}=2.2$.

The potential remains Newtonian, despite the fact that the final
stages of the inspiral we simulate here are close to the innermost
stable circular orbit (ISCO) of the central SMBH ($r_{\rm isco}=3r_{\rm S}$ for a non-spinning BH). 
In reality the disc dynamics would be altered by relativistic effects,
but we chose to start with the simpler, scale-free case before
investigating more realistic additional physics in future work.

In a similar vein, throughout the inspiral we measure the torques
exerted on the BH, but the BH does not respond to these torques. 
Restricting the BH to adhere to an imposed circular orbit is artificial: in reality, gas bound to the BH would imprint a time dependent torque on it.  As an analogy, the torque exerted by the Moon on the Earth is balanced by the torque from the Earth on the Moon.  While the Earth would follow an epicyclic motion, the total angular momentum of the Earth-Moon system cannot change.  One can therefore wonder whether in our case, by prescribing the Earth's orbit, we introduce spurious torques from the Moon and violate angular momentum conservation.  However, one can show using the restricted 3-body approximation \citep{MD2000} in which the Moon's mass is zero (consistent with the massless disc assumed in our simulation runs) that the Moon's orbit is has a front-to-back symmetry around the Earth, spending equal times ahead and behind the Earth, and the net orbit-averaged torque vanishes.
Not implementing a `live' BH allows us to observe gas asymmetry and makes a single calculation scalable to any
value of $\Sigma_0$.  As we show below, the migration rate of the BH
is strongly dominated by GW emission, and gas torques only impart a
very small deviation.

\subsection{Accretion Prescription}

To model accretion onto the migrating secondary BH ($M_2$), we implement a sink
prescription similar to \citet{Farris2014} and
\citet{Tang2017}. Within a sink radius centered on the BH, the surface
density of the gas is reduced on the local viscous timescale, assuming the gas surrounding the secondary settles into a mini-disc with parameters $\mach=20$ and $\alpha = 0.03$. The sink
radius is set to be the smoothing length of the gravitational
potential $\epsilon$, since we do not resolve the BH's event
horizon. There are 15 cells across the accretion sink, which in
physical units extends to 125 Schwarzschild radii (in radius) of the
secondary BH.

The accretion timescale for the secondary BH is given by the viscous time of the mini-disc,
$t_{\rm  visc}(\epsilon) = ({2}/{3}) ({\epsilon^2}/{\nu})$.  This is
implemented with a new source term, which reduces the surface density
inside the sink at the rate
\be
\label{eq:sink}
\frac{d \Sigma}{dt} = -\frac{\Sigma}{t_{\rm visc}} \exp{-(r_2/r_{\rm sink})^4}.
\ee
The exponential factor acts to smooth the sink radius boundary, in
order to reduce numerical artifacts that arise from discontinuous
changes in the surface density. 

 In the context of a steady $\alpha$-disc model, choosing a Mach number 
together with physical units for length and surface
density implies an accretion rate which tends to be well in excess
of the Eddington rate ($\dot{M}_{\rm Edd} \equiv L_{\rm Edd}/(\epsilon_{\rm eff} c^2$), where $L_{\rm Edd} = 4 \pi G c \kappa^{-1}_{\rm es} M$ is the Eddington luminosity, $\kappa_{\rm es}$ is the electron scattering opacity, and $\epsilon_{\rm eff}$ is the radiative efficiency).  Likewise, the accretion rate
onto the secondary (Eq. 6) with our choice of $M = 20, \alpha = 0.03$, and
our chosen length units implies $\dot{M}/\dot{M}_{\rm Edd}>1$ for $\Sigma_0>10 \, \rm g\,  cm^{-2}$.
Our estimates for $\Sigma_0$, derived in Section~\ref{sec:results}, exceed this limit and imply an accretion rate that is $10^3 - 10^8 \, \dot{M}_{\rm Edd}$. 

In other words, we are simulating discs that are unrealistically thick and
have unrealistically high accretion rates.  Unfortunately, we, as well as
all similar global numerical disc studies, are unable to model discs that are
sufficiently thin to correspond to sub-Eddington accretion rates. Such thin, cool discs
are numerically challenging to simulate.  In addition, our simulations neglect radiation,
which is inconsistent with the large luminosity expected from near- or super-Eddington
accretion. Radiation would certainly play a role in the gas properties around the secondary
at these high accretion rates (although, we note that radiation-hydrodynamical simulations
by \citealt{JSD2014} of super-Eddington accretion discs find that radiation preferentially
escapes the disc in the vertical direction, allowing radiatively efficient accretion
close to that expected in a thin disc).

For simplicity, we neglect the impact of radiation in the present work. 
 In reality the accretion efficiency
onto the secondary BH is uncertain, and our fiducial choice is an
estimate.  Future work will explore the dependence of the gas dynamics
and torques on the accretion efficiency.

\begin{table*}
\begin{tabular}{llll}
\specialrule{0.8pt}{1pt}{1pt}
 & \hspace{38mm} \sc Simulation Parameters &  \\
\hline\midrule
\vspace{-0.4cm}
\\
\multicolumn{1}{l}{ Parameters} &
\multicolumn{1}{l}{} &
\multicolumn{1}{l}{[cgs]}&
\multicolumn{1}{l}{[code]} \\
\specialrule{0.5pt}{1pt}{1pt}
\vspace{-0.3cm}
\\
  $\alpha$                  &  Viscosity parameter    &  $0.03$     & $0.03$     \\ \specialrule{0.2pt}{1pt}{1pt} 
    $\mach$                  & Mach number &   $20$    & $20$     \\ \specialrule{0.2pt}{1pt}{1pt} 
     $\epsilon  $    &Smoothing length  &   $125 \, r_{\rm S}(\rm M_2) = 3.7 \times 10^{10} \, cm$    & 0.025     \\ \specialrule{0.2pt}{1pt}{1pt} 
          $ r_{\rm sink}$    & Sink radius   &    $125 \, r_{\rm S}(\rm M_2)$    & 0.025     \\ \specialrule{0.2pt}{1pt}{1pt} 

$\rm M_{\rm BH}$    & Mass of primary BH & $10^6 \, \rm M_{\odot}$ & $G{\rm M_{\rm BH}}=1.0$ \\ \specialrule{0.2pt}{1pt}{1pt}
  $\rm q$                  & Mass ratio $\rm M_2/M_1$ & $10^{-3}$      & $10^{-3}$     \\ \specialrule{0.2pt}{1pt}{1pt} 
  $\rm n_{\rm orbits}$ & Total \# of simulated binary orbits  & 9,721 &  9,721 \\ \specialrule{0.2pt}{1pt}{1pt}
  $\rm r_{\rm min}$  &  Final binary separation & 5$ \, r_{\rm S}(\rm M_{1})=1.5 \times 10^{12} \rm\,cm$ &  1.0 \\ \specialrule{0.2pt}{1pt}{1pt}
    Derived  parameters &   &  &   \\ \specialrule{0.5pt}{1pt}{1pt}
  $\rm \dot{r}/r$ & GW inspiral rate at final separation &-2.59 $\times\,10^{-7} \, \rm s^{-1}$ &-4.04 $\times \, 10^{-5}$    \\ \specialrule{0.2pt}{1pt}{1pt}
$\rm r_{\rm max}$ & Initial binary separation & 11$r_{\rm S}(\rm M_{1})=$ 3.3 $\times\,10^{12}\rm\,cm$ & 2.2 \\ \specialrule{0.2pt}{1pt}{1pt}
$\rm t_{\rm total}$  & Total simulation time & 2.16 $\times \,10^{7} \, \rm s$ = 0.68 \,yr & 22,088.3 $\times \, 2\pi$ \\ \specialrule{0.5pt}{1pt}{1pt}
    &\hspace{40mm} \sc LISA-related Parameters &  \\
\hline\midrule
\vspace{-0.4cm}
\\
\vspace{-0.4cm}
\\
$\rm f_{\rm min}$     & Initial GW frequency of binary & $0.626 \, \rm mHz$ &    \\ \specialrule{0.2pt}{1pt}{1pt}
$\rm f_{\rm max}$    & Final GW frequency of binary  & $2.042 \,\rm mHz$ &    \\ \specialrule{0.2pt}{1pt}{1pt}
$z$    & Redshift (for detectability estimate) & $1$ &    \\ \specialrule{0.2pt}{1pt}{1pt}
$\rm \tau$ & LISA mission lifetime & $5$ yrs \\ 
$\rm L$ & LISA arm length & 2.5 million km \\
$\rm N$ & Number of laser links & 6 \\ 
\specialrule{0.8pt}{1pt}{1pt}
\end{tabular}
\caption{Definition of parameters used throughout the paper, with their adopted values in both physical and simulation units.}
\label{table:parameters}
\end{table*}

\section{Simulation results}
\label{sec:results}

Here we define the various components of the torque exerted on the
inspiraling BH before directly comparing them in the simulation.

The dominant mechanism for angular momentum loss of the secondary BH
is GW emission (\S~\ref{sec:GRtorque}), for which the torque is
derived from the quadrupole formula:
\be
\label{eq:Tgw}
T_{\rm GW} = -\frac{1}{2} M_2 r \dot{r}_{\rm GW} \Omega_2.
\ee
Note that we are assuming the center of mass is at the position of the
primary, $\Omega_2$ is the Keplerian orbital frequency of the secondary,
and we quote the torques exerted on the secondary.

Gas can impart a torque on the secondary in two different ways -- by
gravitational interaction and by accretion. The gravitational torque
$T_{\rm g}$ arises from the gravitational force exerted by the gas.
We calculate this torque by summing up the $\phi-$component of the gravitational
force ${\bf g}_{\phi}$ crossed with the binary lever arm ${\bf r}$
over all the grid cells in the disc,
\be
\label{eq:Tg}
T_{\rm g} =
\sum | {\bf g}_{\phi} \times {\bf r} |
\ee
where $|\bf r|$ is the separation of the binary. We describe this torque in
detail in \S~\ref{sec:GRtorque} and distinguish between positive and
negative contributions from different regions of the disc.

Accretion onto the BH is another mechanism for angular momentum
gain/loss. Assuming that the relative linear momentum of the accreted
gas is added to the BH, the accretion torque $T_{\rm acc}$ is
calculated by summing the relative momenta contributed by the cells
within the sink radius:
\be
\label{eq:Tacc}
T_{\rm acc} = 
\sum_{\rm sink} \dot{m} | {\bf v_{\rm rel}}\times {\bf r} |,
\ee
where
${\bf v_{\rm rel}} = {\bf v_i} - {\bf v_{\rm BH}}$
is the velocity of the gas relative to the BH,
and $\dot{m}$ is the accretion rate inside the sink, i.e. the integral
of equation~(\ref{eq:sink}).  As discussed in \S~\ref{sec:acctorque}
below, we find this component of the torque to be relatively
insignificant.

Analytical estimates for Type II migration torques often utilise the
viscous torque as a reference, which follows directly from
equation~(\ref{eq:tvisc}) and is given by
\be
\label{eq:Tnu}
T_{\nu} = -3 \pi r^2 \Omega_2 \nu \Sigma.
\ee

The magnitude of gas torques
(equations~\ref{eq:Tg},~\ref{eq:Tacc},~and~\ref{eq:Tnu}) all depend
linearly on the normalisation of the surface density, a parameter we
are free to choose. Estimates for $\Sigma$ vary by several orders of
magnitude depending on the chosen accretion disc model. In a
steady-state Shakura-Sunyaev disc (\citealt{SS1976}), $\Sigma$ is
determined primarily by the assumed accretion rate $\dot{M}$ and the
viscosity parameter $\alpha$. In the inner regions of accretion discs
where radiation pressure becomes dominant, $\Sigma$ is heavily
dependent on whether viscosity scales with the gas pressure or with
the total pressure (including radiation).  
 For the system we consider here, the inspiraling BH is deep within the radiation-pressure dominated zone.
We adopt two estimates for our
normalization $\Sigma_0$ (at the secondary's final radius at $r_{\rm
  min}=5 r_{\rm S}$) for the inner regions of thin, near-Eddington
accretion discs that utilise each of these assumptions, representing
low-- and high--end estimates which bracket the range of expected
densities.

We normalise our disc densities to represent AGN accreting at near-Eddington rates
with a radiative efficiency  $\epsilon_{\rm eff}=0.1$.
The low estimate is obtained from the seminal model for a thin,
viscous accretion disc by Shakura-Sunyaev (i.e. $\alpha$-disc;
\citealt{SS1973}), in which the viscosity is proportional to the total
(gas + radiation) pressure. For this model, the surface density in the
radiation-pressure dominated inner region is given by
\be
  \label{eq:Sigma_alpha}
\Sigma_{\alpha} = 88.39 \,  \left(
\frac{\alpha}{0.03} \right)^{-1} \! \left( \frac{\dot{M}}{0.1
  \dot{M}_{\rm Edd}} \right)^{-1} \!\left( \frac{r}{5 r_{\rm S}}
\right)^{3/2}
\, {\rm g \, cm^{-2}}
\ee
where the fiducial choice for the accretion rate is $10\%$ of the
Eddington rate.  In case the viscosity is proportional only to the gas
pressure (i.e. for a so-called $\beta$-disc), the surface density at
the same accretion rate is much higher.  We estimate the surface
density in this second model (see \citealt{Haiman+2009}) as
\begin{multline}
  \label{eq:Sigma_beta}
  \Sigma_{\beta} = 1.55 \times 10^7 \, \left( \frac{\alpha}{0.03} \right)^{-4/5} \! \left( \frac{\dot{M}}{0.1 \dot{M}_{\rm Edd}} \right)^{3/5} \times \\
\times \left( \frac{M}{10^6 M_{\odot}} \right)^{1/5} \! \left( \frac{r}{5 r_{\rm S}} \right)^{-3/5}
\,{\rm g \, cm^{-2}}.
\end{multline}

\begin{figure}
\begin{center}
\includegraphics[width=.5\textwidth]{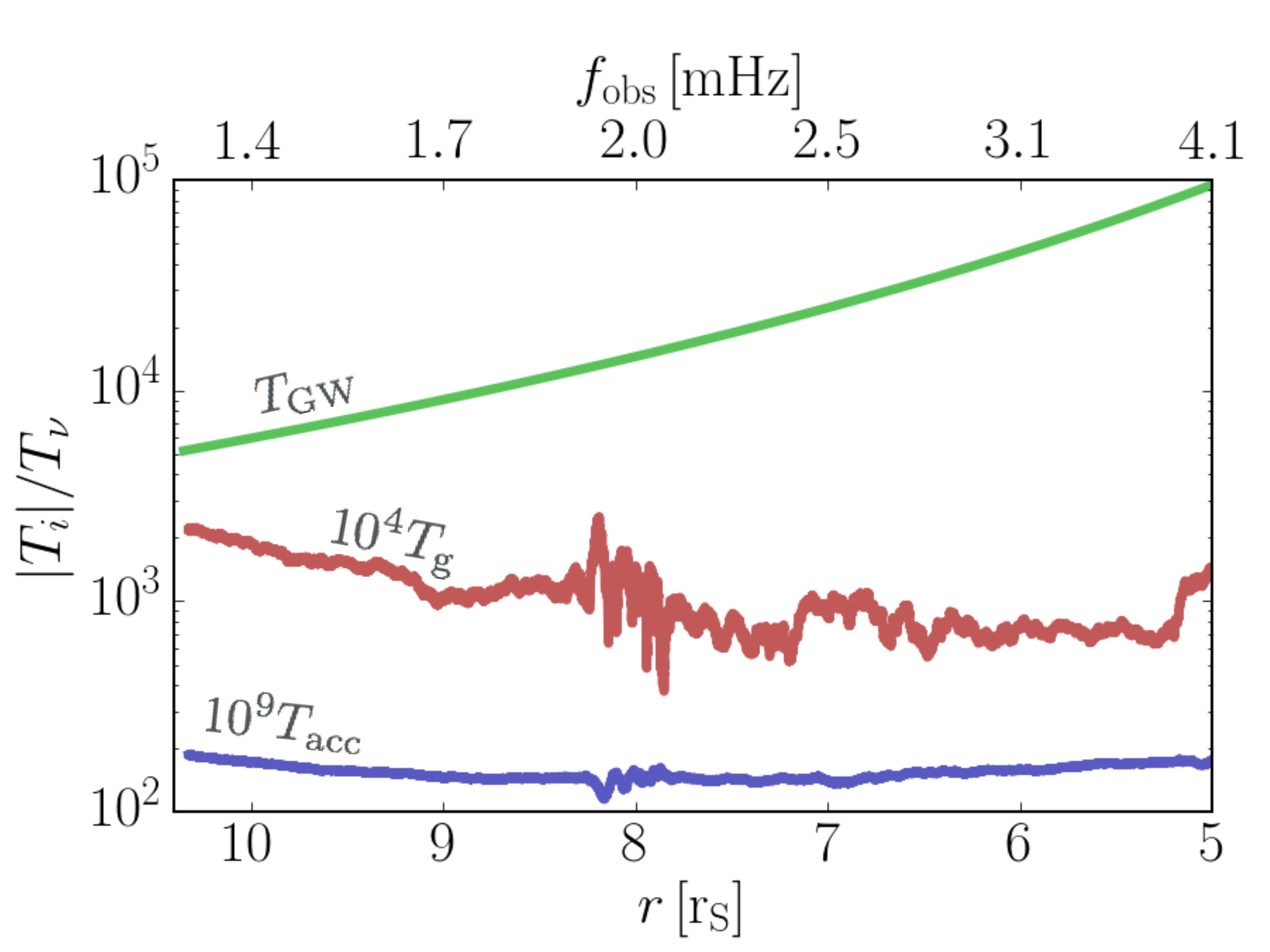} 
\caption{Different components of the total torque exerted on the
  secondary BH: gravitational torque ($T_{\rm g}$) and accretion
  torque ($T_{\rm acc}$) from the gas disc, as well as the torque from
  GW emission ($T_{\rm GW}$).  All three torques are scaled by the
  fiducial viscous torque at the secondary's location. The gas torques
  are normalised to correspond to a $\beta$-disc (see
  equation~\ref{eq:Sigma_beta}).  Note the overall different scale
  for each torque shown.  The GW and accretion torques are both inward,
  while the gravitational torque is outward.}
\label{fig:GWtorq}
\end{center}
\end{figure}

\subsection{Gas torques versus the GW torque}
\label{sec:GRtorque}

The gas torques measured in the simulation, along with the GW
torques, are compared in Fig.~\ref{fig:GWtorq}. All torques are
shown scaled by the viscous torque and as a function of the evolving
binary separation.  The first point we address is the strength of the
gas torques relative to the GW torque.
The gas torque we measure differs from the analytical estimates (based
on the viscous torque $T_{\nu}$) in both magnitude and direction. As
Fig.~\ref{fig:GWtorq} shows, the torque we find is $5-10$ times
weaker than $T_{\nu}$.

 As Figure~\ref{fig:GWtorq} shows, the gravitational torque we find is $5-10 \times$ weaker than the fiducial viscous torque.   Note that for gap-opening planets on fixed circular orbits, the standard Type-II torque would match the viscous torque $T_\nu$ to within a factor of few.  This torque would cause the planet to migrate inward with a drift velocity matching that of the gas, although when the disc mass is below the planet mass, angular momentum conservation implies that the drift velocity would decrease linearly with the ratio $4\pi r^2 \Sigma_0/M_2$ \citep{SC1995}.

\begin{figure}
\begin{center}
\includegraphics[width=.5\textwidth]{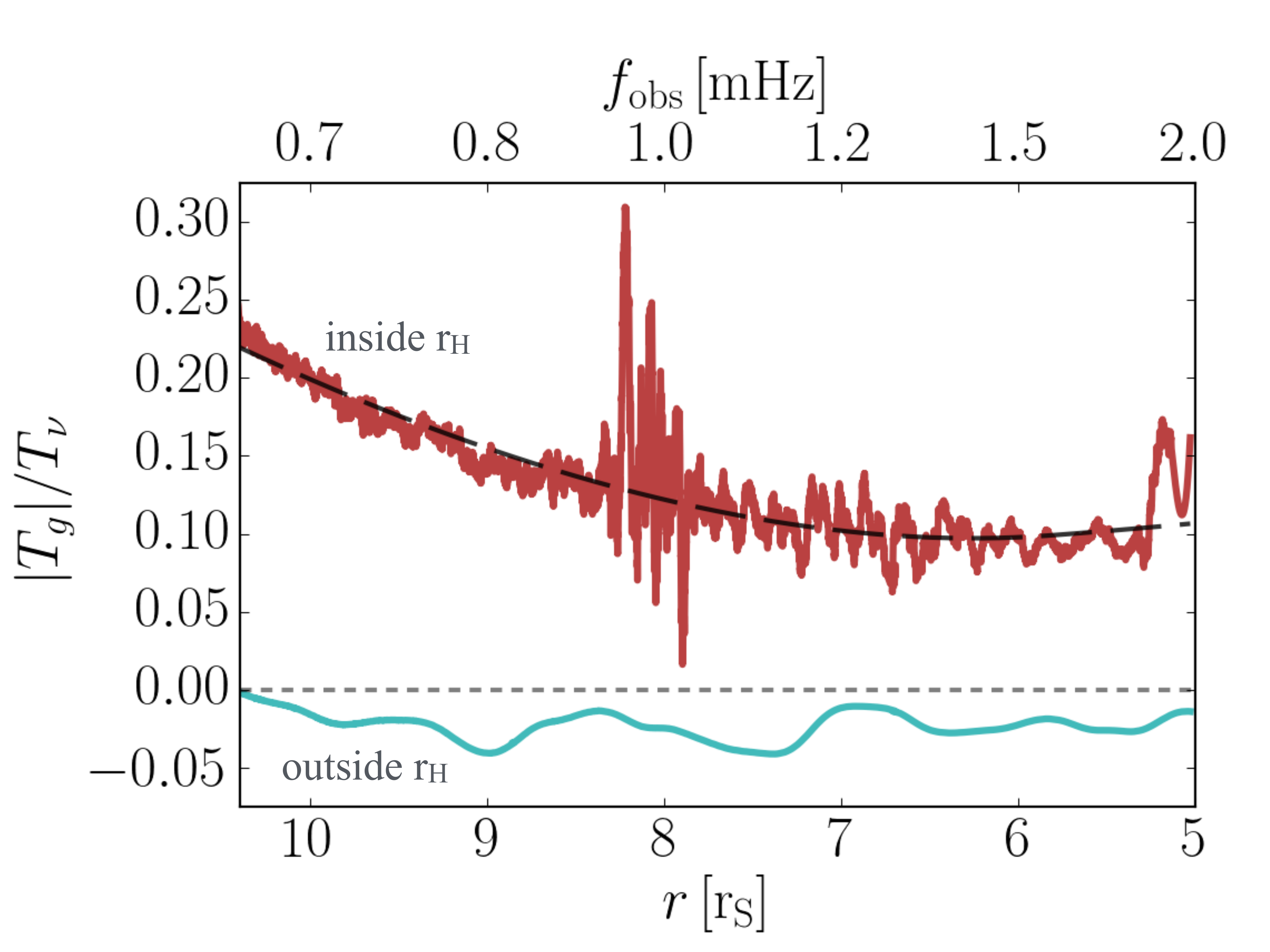} 
\caption{Gravitational torque $T_{\rm g}$ exerted by different regions
  of the gas disc onto the secondary BH, as a function of binary
  separation.  The red (upper) curve shows torques from within the
  Hill sphere, and the green (lower) curve shows torques from outside
  this region. Both torques are scaled by the viscous torque. The
  dashed curve shows fitting formula we adopt for our LISA SNR
  computations (\S~\ref{sec:detectability}).}
\label{fig:torq_ex}
\end{center}
\end{figure}

In addition to being an order of magnitude weaker than standard Type II torques, for our disc parameters, the gravitational torque causes the secondary the migrate {\it outward} rather than inward
\citep[outward torques in isothermal discs have also been noted
  by][]{Duffell2015}. The torque also changes with migration rate,
decreasing in strength as the secondary BH moves inward and
accelerates.  For the case that the surface density is high, resulting
in the strongest torque (in the $\beta$-disc model), the gas torque is
still several orders of magnitude weaker than the GW torque
(Fig.~\ref{fig:torq_ex}). Nevertheless migration can still produce a
detectable deviation in the GW signal, as we discuss in~\S~\ref{sec:detectability}.

Our simulation results can be used to understand the origin of gas
torques in more detail, and how the different components depend on the
accretion or migration rate of the BH. The entirety of the disc exerts
some gravitational force on the secondary BH, including both the inner
and outer disc, streams across the gap, and gas near the BH itself. We
include all of these regions to calculate the total gas torque.

We find that gas closest to the BH, particularly within the Hill
sphere, is the dominant contributor to the torque due to its close
proximity. The Hill sphere is an approximation of an embedded body's
sphere of influence in the presence of a more massive body at a
distance $a$. Its radius is estimated as
\be
r_{\rm H} = \left( \frac{q}{3} \right)^{1/3} a.
\ee
As gas flows across the gap, it can continue to replenish the net
angular momentum of material in the Hill sphere of the secondary.
This can cause an asymmetry within the Hill sphere, and a slight
increase in the gas density upstream from the BH, leading to a
consistently positive (outward) torque.  This torque has been seen in
other works (e.g. \citealt{DAngelo2005}).  Accurately capturing the
gas morphology within the Hill sphere requires high spatial resolution.
Further work investigating the gas streamlines, particularly with
high-resolution 3-dimensional simulations, will likely be necessary to
accurately measure the torque arising from this asymmetry, and to
fully understand the reason for the small front-back asymmetry that
yields the net torque.  
  
We note that in disc-satellite calculations, the torque within the
Hill sphere is often ignored or ``damped'', based on the assumption
that material within the Hill sphere is bound to the perturber
(e.g. \citealt{ValBorro2006, Durmann2017}).  
 However, gas flows through the Hill sphere $-$ while
 some of it may become bound or accreted, certainly the majority 
flows across the gap to supply the inner disc. 
Asymmetrically
distributed gas in this region can exert a net torque on the BH, which
must be included if the BH and the gas is separately resolved and
followed. 
This asymmetry, unlike a bound `orbiter', is constantly supplied as gas flows across the gap. In our case the accumulation of gas leading the BH's orbit occurs because gas preferentially flows ahead of the BH and/or because it slows down (and hence spends more time) in this region. 

 Further evidence for torque inside the Hill sphere is present in simulations by \citet{Crida2009}, in which
a live BH experiences changes in migration rate when torques within the Hill sphere are truncated. In their case the planet migrates inward more quickly when the Hill region is included, suggesting an accumulation of gas in the trailing side of the planet's orbit. This is likely due to a difference in disc parameters since the dynamics are sensitive to viscosity and mach number. 
Ultimately truncating the torque in the Hill sphere gives a poor approximation to a self-consistent, highly resolved torque calculation, which suggests that this region is important. 
 Our simulations sufficiently resolve
gas flow within the Hill sphere, and thus we choose to not omit this
region when computing the total torque.

\begin{figure*}
\begin{center}
\begin{tabular}{cc}
\includegraphics[width=1.\textwidth]{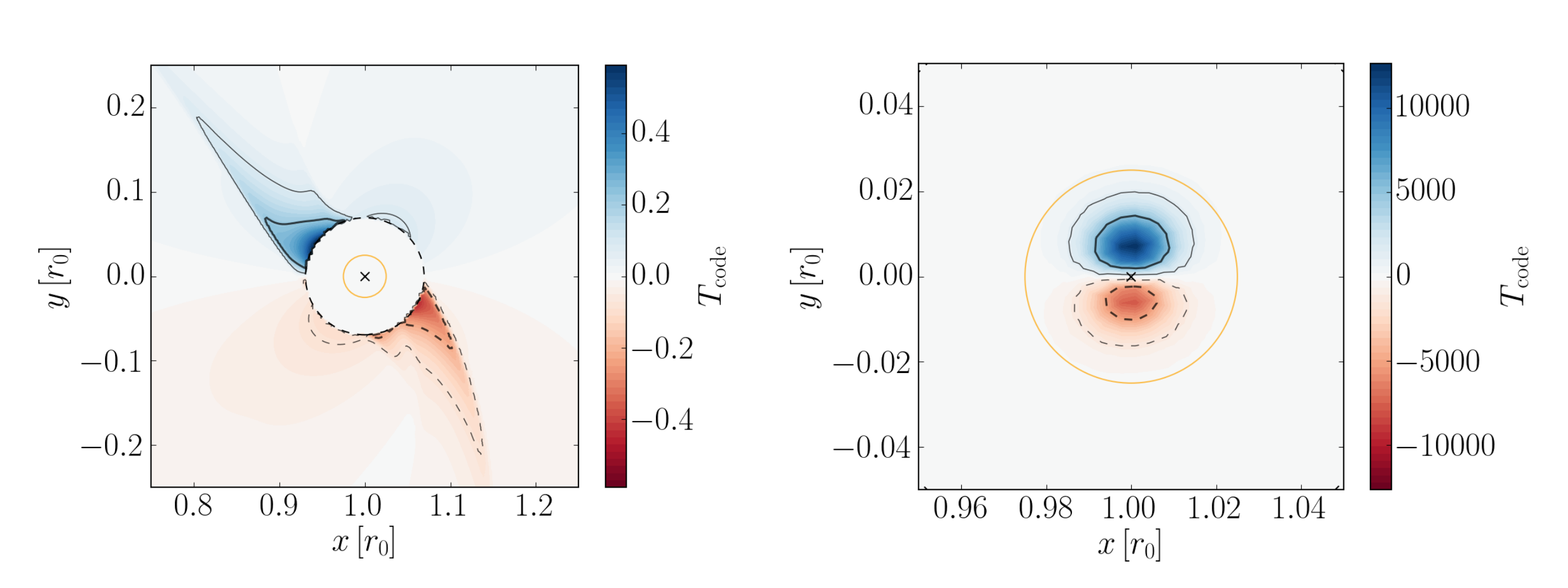}
\end{tabular}
\caption{2D contours of torque surface density, comparing the torques
  contributed by gas within the Hill sphere (right panel) to torques
  from gas outside this region (left panel). }
\label{fig:torqcont}
\end{center}
\end{figure*}

Fig.~\ref{fig:torq_ex} shows the torque broken into two components
-- inside and outside the secondary's Hill sphere -- along with smooth
fits to the data which we use for calculating detectability in
\S~\ref{sec:detectability}.  The Hill sphere torque is qualitatively
different in that it settles to a positive value (it pushes outward). 
The gas from elsewhere in the disc, including the
inner and outer discs, as well as the streams connecting the two
regions, exerts a negative (inward) torque that is 1-5\% of the
viscous torque.  Fig.~\ref{fig:torqcont} shows 2D torque density
contours of each of these components. The left panel illustrates the
torque density outside of the Hill sphere, which is dominated by the
streams flowing across the secondary BH's orbit. The right panel zooms
in on the torque density inside the Hill sphere, showing that there is
a slight density increase in the gas in front of the BH (for a
counterclockwise orbit). This asymmetry is responsible for the
positive torque.

\begin{figure}
\raggedleft
\includegraphics[width=.5\textwidth]{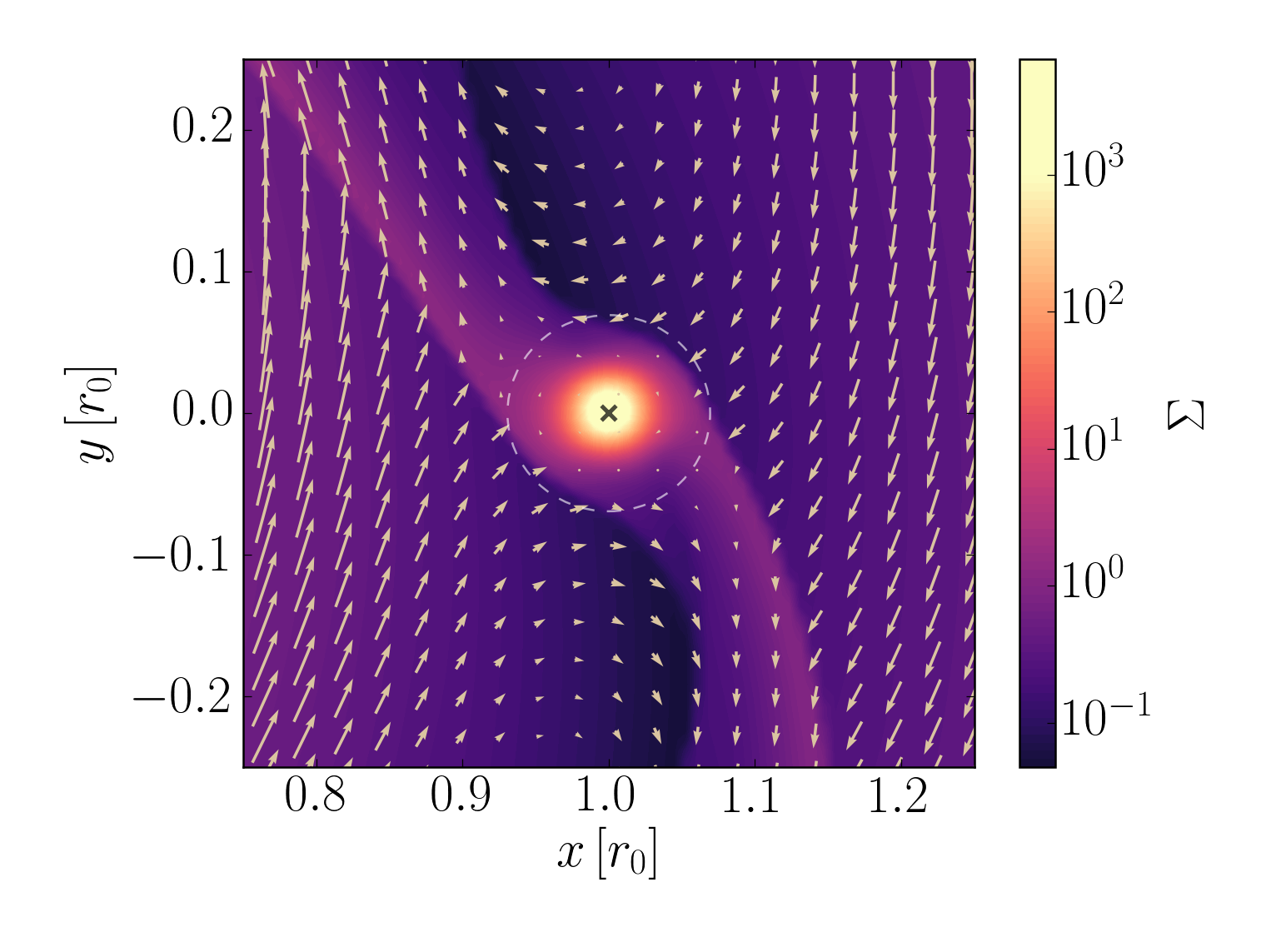} 
\caption{Velocity field in the frame co-rotating with the binary,
  overlaid on the surface density contour. Low relative velocities
  around the secondary BH indicate the build-up of a quasi-stationary
  atmosphere, rather than a near-Keplerian mini-disc.}
\label{fig:velocity}
\end{figure}

\subsection{Accretion torque}
\label{sec:acctorque}

For the fiducial accretion rate, we find that the accretion torque is
negligible compared to the gravitational torque from the gas -- their
magnitudes differ by $\sim6$ orders of magnitude, as seen in
Fig.~\ref{fig:GWtorq}. This is because the relative velocity of the
gas near the black hole is effectively negligible. We show this in
Fig.~\ref{fig:velocity}, with a snapshot of the velocity field in a
frame co-rotating with the binary. The gas close to the BH has low
angular momentum and resembles a quasi-stationary atmosphere, rather
than a near-Keplerian mini-disc.

 Accretion plays a minimal role in our simulations. 
The most conspicuous effect of accretion is to reduce the density of
the gas near the BH, which leads to a decrease in the positive
component of the torque. 
 Without including a sink, our results 
show the same asymmetry within the Hill sphere, albeit with more gas (and a larger positive component of the torque). The torque from elsewhere in the disc is unaffected by our sink prescription. 
We hypothesize that more efficient accretion (as well as
feedback)
would lead to more depleted gas density and less positive torque.
A drastic increase in accretion rate may lead to other differences, since it would
steal gas that would otherwise flow to the inner disc.  We leave an
investigation of the dependence of the torques on the accretion rate
in the $q\neq 1$, GW-driven case to future work.

\section{Detectability of gas imprint by LISA}
\label{sec:detectability}

\subsection{Drift in the accumulated GW phase}

In this section, we estimate the deviation from the vacuum GW signal
caused by the gas disc torques, and assess its detectability by LISA.

The total accumulated phase of a gravitational wave event can be
obtained by integrating over the total frequency evolution
\be
\phi_{\rm tot} = \int_{t_0}^{t_0 + t_{\rm obs}}\! \dot{\phi}_{\rm GW} \, {\rm dt} = 2 \pi \int_{t(f_{\rm min})}^{t(f_{\rm max})} f_{\rm GW} \, {\rm dt}
\ee
where $t_0$ is an arbitrary reference time when the LISA observation
begins, $t_{\rm obs}$ is the total observation time, $f_{\rm min}$ and
$f_{\rm max}$ bracket the corresponding observed frequency range,
$f_{\rm GW}(t) = \Omega (t)/\pi$ is the GW frequency, which is twice
the binary's orbital frequency $\Omega (t)$, and $\phi_{\rm GW}$ is in radians.
Assuming the orbit remains circular throughout the inspiral, changing
the integration variable to orbital separation $r$ gives
\be
\phi_{\rm tot} = -2\pi
\int_{r_{\rm min}}^{r_{\rm max}}\! \frac{f_{\rm GW}}{\dot{r}} \, {\rm dr},
\ee
where $\dot{r}$ is the (negative) radial inspiral velocity corresponding to the
angular momentum evolution of the binary.  In our case, both GW
emission and gas torques change the angular momentum, so
the net evolution can be described by the sum of both components
\be
\label{eq:ldot}
\dot{r} = \dot{r}_{\rm GW} + \dot{r}_{\rm gas},
\ee
and the accumulated phase is given by
\be
 \phi_{\rm tot}  = -2\pi \int_{r_{\rm min}}^{r_{\rm max}}\! \frac{f_{\rm GW}}{\dot{r}_{\rm GW} + \dot{r}_{\rm gas}} \, {\rm dr}.
\ee
Because the effect of gas is much smaller than GWs ($\dot{r}_{\rm gas}
\ll \dot{r}_{\rm GW}$) the difference between the accumulated phase
with and without gas,
$\delta \phi \equiv \phi_{\rm  GW+gas} -  \phi_{\rm GW}$,
can be simplified as follows:
\be
\label{eq:delphi}
\delta \phi = 2 \pi
\int_{r_{\rm min}}^{r_{\rm max}}\! \frac{f_{\rm GW} \, \dot{r}_{\rm gas}} {\dot{r}^2_{\rm GW}} \left [1+\mathcal{O}\left( \frac{\dot{r}_{\rm gas}}{\dot{r}_{\rm GW}}  \right)^2 \right] \, {\rm dr}
\ee
Our simulation provides a direct measurement of $\dot{r}_{\rm gas}(\Sigma_0,r)$.  We simplify the gravitational torque with a fit to the numerically measured value from the simulation. 
The fit for the total torque (adding the components inside and outside of the Hill sphere) has the form
\be
\label{eq:fit}
T_{\rm fit} = \Sigma_0 (A r^2 + Br + C)
\ee
with $A = 1.58 \times 10^{16} \rm \, cm^2 \, s^{-2} $, 
$B = -5.44 \times 10^{28} \rm \, cm^3 \, s^{-2} $, and 
$C = 5.76 \times 10^{40} \rm \, cm^4 \, s^{-2}$, 
This fitting formula is shown together with the numerically measured
torques in Fig.~\ref{fig:torq_ex}. For $r<6r_{\rm S}$, the spike at
the end of the simulation is likely numerical, so in the range
$3 r_{\rm S} \leq r \leq 6 r_{\rm S}$, we set the torque to remain a constant
(we assume $3 r_{\rm S}$ represents the end of the inspiral phase).

Using this fit for the total gas torque in
equation~(\ref{eq:delphi})
with the relation $\dot{r}_{\rm gas} = 2 \dot{\ell}_{\rm gas}
r^{1/2}(GM)^{-1/2}$ (where $\dot{\ell}_{\rm gas}=T_{\rm gas}/M_2$
is the rate of change of the secondary's specific angular momentum),
we compute the phase drift $\delta \phi$ over an observed frequency
window from $f_{\rm min}$ to $f_{\rm max}$.  Unless stated otherwise,
we use the total gas torque, including the contribution inside the
Hill sphere.

\subsection{Signal to noise ratio of the waveform deviation}

\begin{figure}
\begin{center}
\includegraphics[width=.5\textwidth]{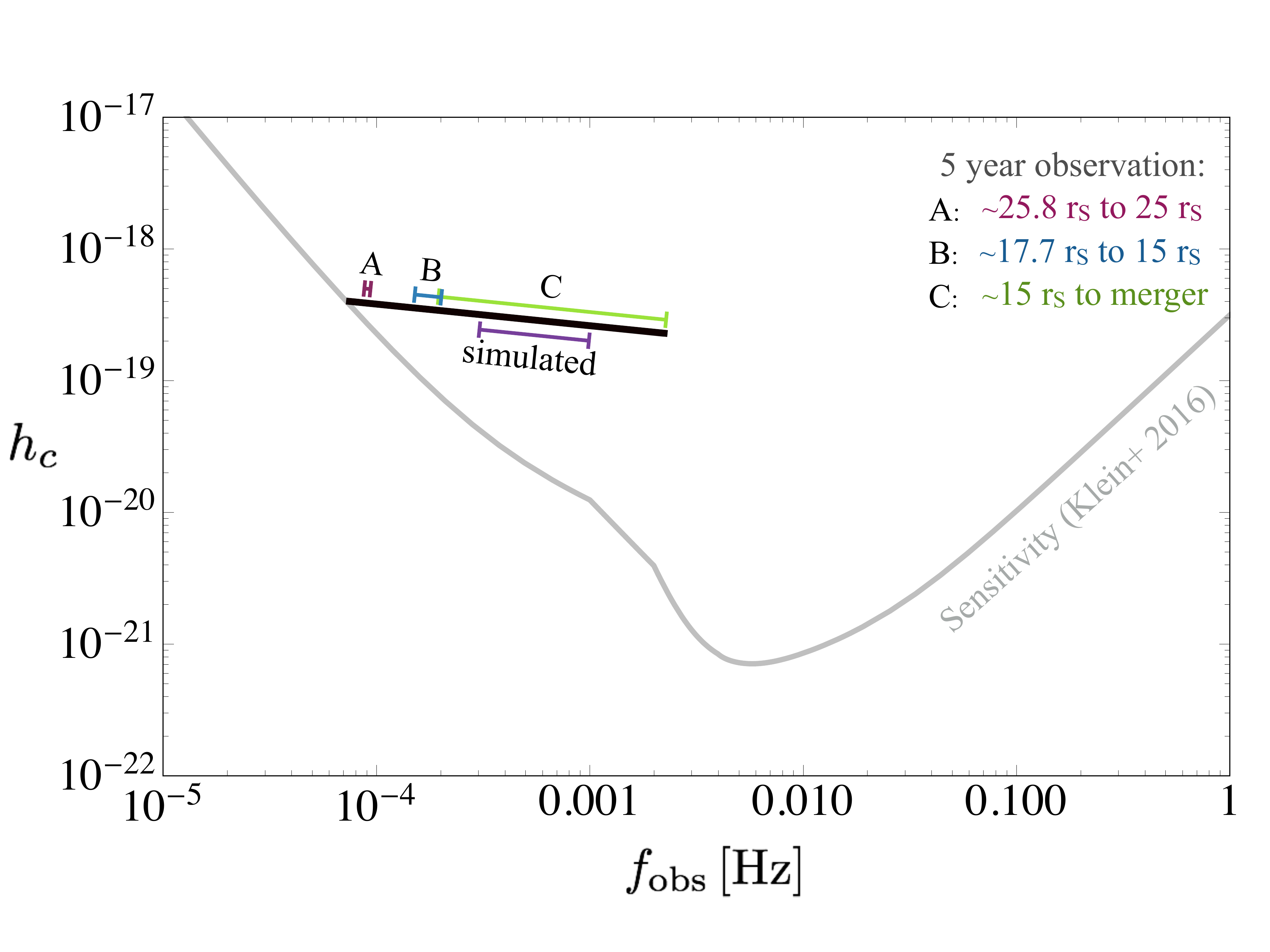} 
\caption{The characteristic strain amplitude of the binary inspiral,
  for an $M_1=10^6 M_{\odot}$, $M_2=10^3 M_{\odot}$ IMRI at redshift
  $z=1$, as a function of the observed GW frequency. The top bracketed regions delineate three different 5--year
  observation windows as labeled, and the bottom bracket highlights
  the portion we cover in our simulation ($\sim\!0.7 $yr in the binary rest frame).}
\label{fig:strain}
\end{center}
\end{figure}

To estimate the detectability of deviations from the vacuum inspiral
waveform, we compute the signal--to--noise ratio (SNR) of the
deviation produced by gas, compared to an event occurring in vacuum.
The detectability of the gas-induced phase drift in an event depends
on the strength of the torque (and thus linearly on the disc mass) as
well as the frequency range of the LISA observation, since this
determines the loudness of the event (the strain amplitude $h$),
the number of observed orbits during which the phase
drift can accumulate, and the instrumental noise.

The sky- and polsrization-averaged GW strain amplitude of a source at
comoving coordinate distance $r(z)$ is
\be
h = \frac{8 \pi^{2/3}}{10^{1/2}} \frac{G^{5/3} \mach_c^{5/3}} {c^4 r(z)} f_r^{2/3},
\ee
where $\mach_c = M_1^{3/5} M_2^{3/5} / (M_1+M_2)^{1/5}$ is the chirp
mass and $f_r = f(1+z)$ is the rest-frame GW frequency
(e.g. \citealt{Sesana2005}).

The \emph{characteristic} strain $h_c$ of a periodic source takes into
account the total observation time $\tau$ (i.e. the LISA mission
lifetime) as well as the characteristic number of cycles the source
spends in each frequency band, $n \equiv f^2/\dot{f}$. For
illustrative purposes, in Fig.~\ref{fig:strain} we plot the
characteristic strain of an IMRI
with our chosen mass ratio $q=10^{-3}$, primary mass $M_1 = 10^6
M_{\odot}$ and redshift $z=1$. The top bracketed regions highlight the
evolution of the binary during three different possible 5-year long
observational periods. The bottom bracketed portion delineates the
evolutionary track we cover in our simulation, which corresponds to
$\sim0.7$ yr in its rest frame, or 1.4 yrs if the binary is at $z=1$. The torques outside this regime are extrapolated
using the fitting formula (eq.~\ref{eq:fit}).

Following \citet{Kocsis2011}, the phase drift $\delta \phi$ can be
expressed by the strain \emph{deviation} $\delta \tilde{h}$ in Fourier
space.  If the Fourier amplitude strain of a vacuum waveform is
$\tilde{h}$, the difference in strain due to the phase drift is given
by
\be
\delta \tilde{h} = \tilde{h}(1 - e^{i \delta \phi}),
\ee
assuming the gas-impacted and vacuum waveforms differ only in phase.
The SNR of the deviation $\delta\rho$ can be computed in
Fourier space as
\be
\label{eq:relativeSNR}
(\delta\rho)^2 = 2 \times 4 \int_{f_{\rm min}}^{f_{\rm max}} df \frac{|\delta \tilde{h} (f)|^2}{S_n^2(f) f^2},
\ee
where $S_n(f)$ is the LISA sensitivity per frequency bin taken from
\citet{Klein2016}. The factor of $4$ comes from the normalization of
the one-sided spectral noise density, and the extra factor of 2 arises
from the currently proposed configuration of LISA having 6 links, or
effectively two interferometers.
Similarly, the total SNR of the event is
\be
\label{eq:totalSNR}
\rho^2 = 2 \times 4 \int_{f_{\rm min}}^{f_{\rm max}} df \frac{|\tilde{h} (f)|^2}{S_n^2(f) f^2},
\ee

\begin{figure}
  \includegraphics[width=.5\textwidth]{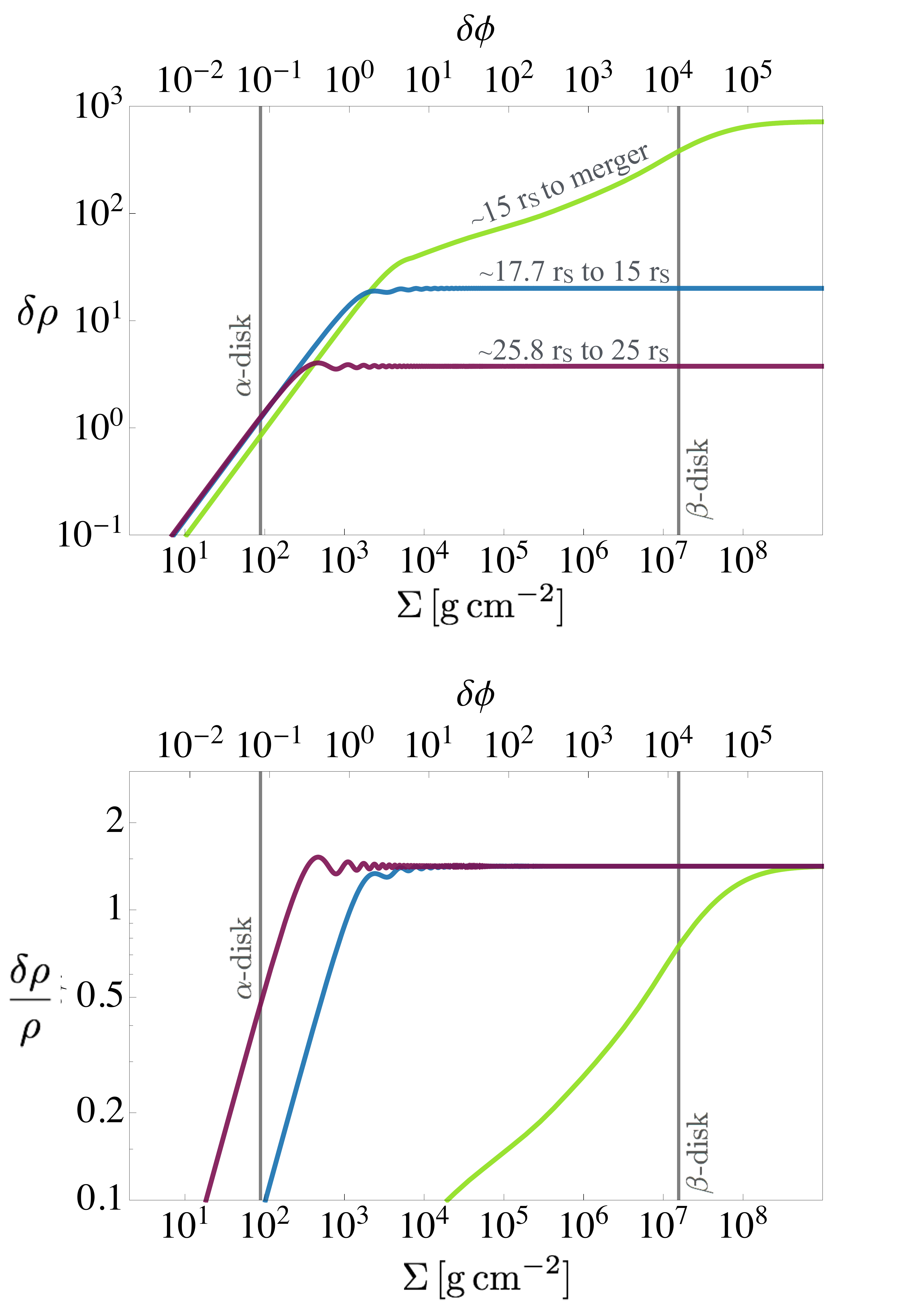}
\caption{The top panel shows the SNR of the gas-induced deviation
  ($\delta\rho$; eq.~\ref{eq:relativeSNR}) in the LISA waveform as a
  function of disc density, for different 5--year observation windows,
  and with our fiducial parameters $M=10^6{\rm M_\odot}$, $q=10^{-3}$,
  and $z=1$.  The lines are labeled with the binary's rest-frame
  separation during the LISA observation, with an assumed total
  observation time of 5 years.  The top axis shows the corresponding
  total accumulated phase drift for each surface density, for the
  (ideal) case of observing the final 5 years to merger (green line).
  The lower panel shows the relative SNR: the SNR of the deviation in
  units of the total SNR of the event ($\delta\rho/\rho$; with $\rho$
  from eq.~\ref{eq:totalSNR}).  }
\label{fig:SNR_sigma_fmin}
\end{figure}

The total inspiral for this particular binary takes $\sim 100$ years 
after the strain enters the LISA frequency band (at $f_{\rm min} \sim
10^{-4}\rm Hz$), so the total SNR depends on what separation the
binary is at (or what GW frequency the binary is emitting) when the
LISA observation begins.  The top panel in
Fig.~\ref{fig:SNR_sigma_fmin} shows the SNR of the gas-induced
deviation as a function of the disc surface density $\Sigma_0$ and the
phase drift $\delta \phi$ for three different 5--year observed
frequency windows. Vertical lines in the figure mark the two estimates
of the surface density for near-Eddington accretion discs
(Eqs.~\ref{eq:Sigma_alpha}~and~\ref{eq:Sigma_beta}). Initially the SNR
scales linearly with surface density, before it saturates around a
particular value once $\delta \phi \approx 2\pi$.

The bottom panel of Fig.~\ref{fig:SNR_sigma_fmin} shows the ratio of
the SNR of the deviation compared to that of the total event,
$\delta\rho/\rho$.  By dividing out the total SNR, this quantity isolates
the impact of the gas.
For the case where the SNR of the deviation is highest (the green
curve in Fig.~\ref{fig:SNR_sigma_fmin}), the SNR never reaches a
saturated value because as the binary chirps towards merger, the
change in the GW frequency results in a considerable contribution to
$(\delta\rho)^2$ from a broad range of higher frequencies.  While the
influence of gas is comparatively stronger than GWs during the earlier
stages of the inspiral, the detectability is less likely because a
smaller frequency range results in a weaker SNR.

The SNR of the deviation (top panel in Fig.~\ref{fig:SNR_sigma_fmin}) is directly relevant to the detectability,
and is dependent both on the strength of the deviations introduced by
the gas, as well as on the total SNR of the event.  While gas effects
are stronger in the earlier stages of the inspiral, the overall SNR
obtained by observing these stages (regardless of any phase change) is
relatively low. If LISA catches the binary $5$ years prior to merger,
however, where its initial rest-frame separation is $r_{\rm min}\sim
10 r_{\rm S}$, it will accumulate a significant amount of SNR as the
binary chirps to merger, largely because the LISA noise is much lower
towards higher frequencies (see Fig.~\ref{fig:strain}).
Thus for this particular binary, the deviation is most detectable in
the final five years of the inspiral.  This would not necessarily be
the case, however, for lower-mass primary BHs, or lower-redshift
events, which would chirp past the minimum of the noise curve.
Adopting the criterion of detectability $\delta\rho \geq 10$, we find
that the gas imprint is detectable for our fiducial binary if the BH
is embedded in gas with surface density $\Sigma_0 \gtrsim 10^{3} \,\rm
g \, cm^{-2}$.

\section{Discussion and Caveats}
\label{sec:discussion}

In the present work we find that, depending on the AGN disc mass, LISA
can detect the migration imprint on gas-embedded IMRIs in the final 5
years of the inspiral.  With our current understanding of AGN, the
typical density of accretion discs is uncertain, and theoretical
estimates range from $\Sigma\sim10^1 - 10^8\, \rm g \, cm^{-2}$ in the
regions of interest ($\lesssim30 r_{\rm S}$). We find that the
deviation is detectable (SNR > 10) if the secondary BH is embedded in
a disc with a surface density $\Sigma \gtrsim 10^{3}\,\rm g
cm^{-2}$. This density is reasonably reached in models of
near-Eddington accretion discs; in particular it is exceeded in
so-called $\beta$-discs.  In our fiducial $\alpha$-disc model, the
torques are too weak to detect.

The difference between $\Sigma$ in these models lies in the physical
mechanism providing the viscosity and its dependence on radiation
pressure, an area of active research in accretion disc dynamics (see \citealt{Blaes2011, YFJ2013}).
The viscosity in $\beta$-discs, in particular, is assumed not to rise
with radiation pressure. As a result, it is much lower than in the
$\alpha$-discs, resulting in a much higher disc surface density at a
fixed accretion rate.  If the signatures of a gas-embedded IMRI are
detected by LISA, then it should be possible to extract information
about the underlying disc,
as even a measurement of a total accumulated phase shift 
will provide (at the least) a lower limit on the disc density.
If both the amplitude and frequency-dependence of the deviation in the GW signal 
are well measured, this will directly probe the density, 
and we expect other disc parameters to be constrained as well 
(e.g. density gradient, temperature, viscosity)
provided the frequency-dependence amongst these parameters is well understood 
and not degenerate. 
Our basic conclusion is that \emph{LISA will have to opportunity to
  probe disc migration physics via gravitational waves}.

We find that the effect of migration torques is stronger during the
earlier stages of the inspiral, but its detectability is highest at
higher frequencies, where LISA is most sensitive, and the
system is chirping rapidly.  For our fiducial binary ($M_1 = 10^6
M_{\odot}$, $q = 10^{-3}$, $z=1$), the gas-induced deviation is most
detectable during the final years of the inspiral, as this is when
LISA accumulates most of the total SNR for the event.
This feature is characteristic of the particular (redshifted) chirp
mass, for which the last several cycles of the coalescence occur at
frequencies near the minimum of the LISA sensitivity curve (see
Fig.~\ref{fig:strain}).  For IMRIs with a higher chirp mass the
merger will occur more quickly, the characteristic frequency is
shifted to lower values, and we expect the contribution to the SNR from
the final stages to be lower. The same may be true for lower chirp
masses, which shift the final few cycles to higher frequencies past
the minimum of the LISA noise curve.  We plan to explore the range of
detectability over various system parameters in future work.

\begin{figure}
\includegraphics[width=.5\textwidth]{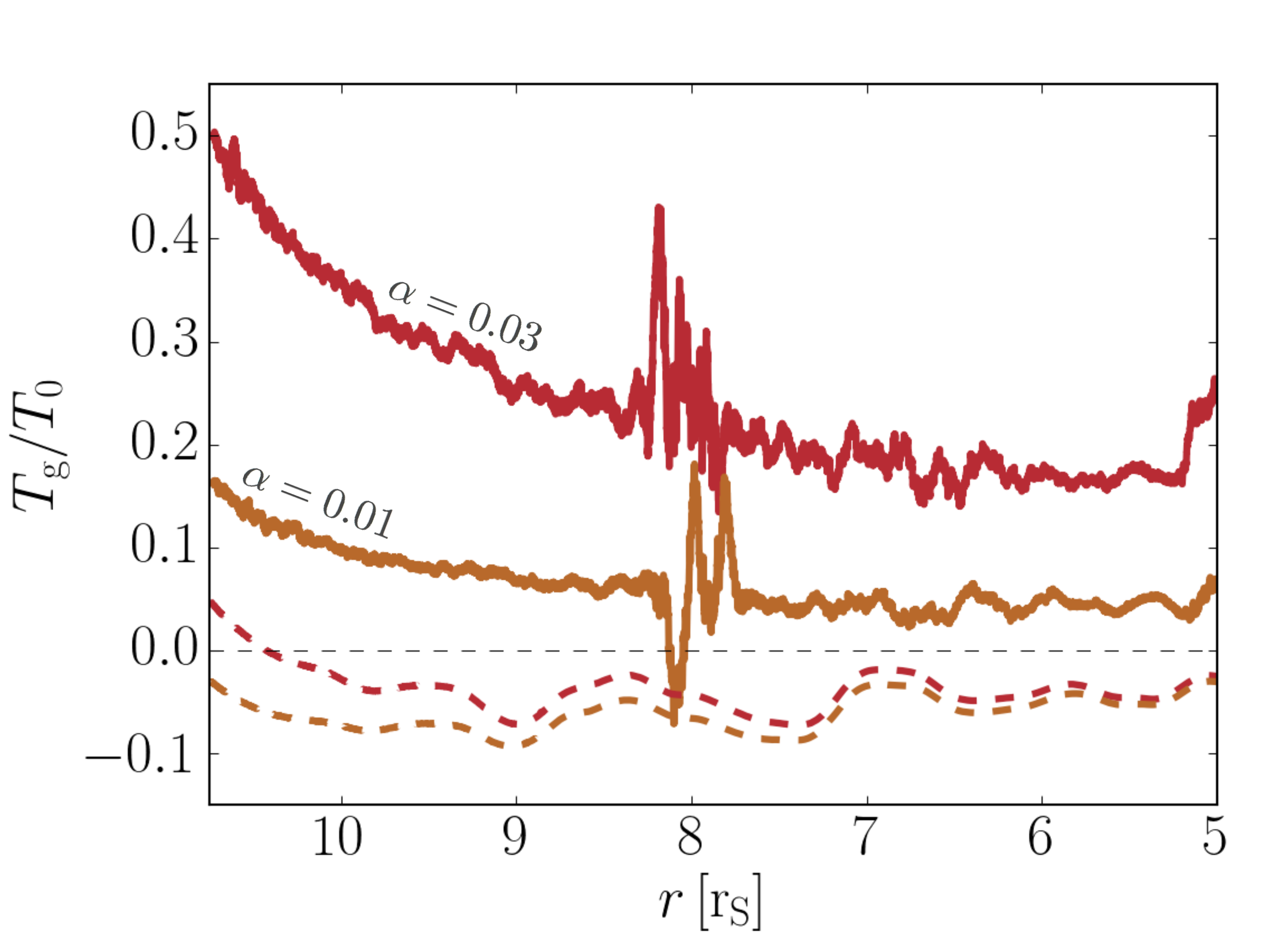}  
\caption{Torques exerted on the secondary BH from gas within the Hill
  sphere (solid curves) and outside the Hill sphere (dashed curves),
  for two runs with different viscosity parameter $\alpha$ as
  labeled. The torque from gas inside the Hill sphere is highly
  sensitive to the viscosity.}
\label{fig:torq_alphas}
\end{figure}

It is worth noting that the torques depend on both disc physics and
the accretion efficiency of the secondary BH. In particular,
simulations of equal--mass binaries find that the build--up of gas
within and near the Hill sphere, and therefore the net torque, scales
with viscosity~\citep{Tang2017}.
To assess the sensitivity in our case, we re-run our fiducial
calculation, except we reduced the viscosity by a factor of three.
The evolution of the gas torques in both cases are compared in
Fig.~\ref{fig:torq_alphas}.
We normalize the torques by the conventional 
Type I torque from \citet{Tanaka2002} provided by $T_0  = \Sigma r^4 \Omega^2 q^2 \mach^2 $, 
which does not depend on viscosity and thus allows for a direct comparison.
As this figure shows, in the new run
with $\alpha = 0.01$, the Hill sphere torque is $3$ times weaker than
in the $\alpha=0.03$ case. On the other hand, the torques from outside
the Hill sphere remain similar, and they tend to converge to the same
value by the end of the inspiral.
This suggests that the increase in the sink timescale
(eq.~\ref{eq:tvisc}) caused by the smaller $\alpha$, is more important
than the change in the dynamics due to the reduced viscosity in the
bulk of the simulation,
although we note that the reduced bulk viscosity may also
play a role in reducing the gas flow around the secondary BH.
We plan to further explore the sensitivity to $\alpha$ and to other parameters in future work.

We also note that in all cases, the migration torque exerted by the
gas outside the Hill sphere is much weaker (by approximately an order
of magnitude) than the torques from inside,
and they also have the opposite sign (resulting in inward migration).
Thus if the torques from inside the Hill sphere were excluded, the phase
drift would be negative, corresponding to a slight increase in the inspiral rate, and the detectability would require
an order of magnitude higher disc surface density.

Our detectability estimates are obtained for an intermediate mass ratio
binary.
For binaries with a more extreme mass-ratio, we expect that the gas
effects would be more easily observable.  First, a lower mass-ratio
binary emits weaker GWs: the GW inspiral time is proportional to
$q^{-1}$ (at fixed total mass and binary separation measured in
gravitational radii).  The total number of cycles observed during the
fixed LISA observation time scales with chirp mass as $\propto
\mach_c^{-5/8}$, yielding more cycles over which the phase drift can
be measured for lower-mass companions.  On the other hand, the scaling
of the disc torques with mass ratio in the GW-driven regime is unknown
and must be computed in future work.  We note here only the
following: If the gas-induced migration time-scale followed the
viscous time, it would be independent of the secondary's mass $M_2$.
However, migration may be slower than the viscous timescale by a
factor that depends on the ratio of the secondary and disc mass,
$M_2/M_{\rm disc}$. In the self-similar models employed in
\citet{Haiman+2009}, the migration timescale is proportional to
$q^{-3/8}$, although simulations by \citet{Duffell2014} demonstrate that 
this dependency may be more complex.
Finally, as the mass ratio decreases and the secondary BH no longer
carves a gap, it will enter the Type I regime,
 in which the torque (provided by $T_0$) scales differently with mass ratio and disc parameters.
As a comparison, if the secondary BH was instead a $10 M_{\odot}$ BH 
(which for the disc parameters
we adopt in this work, puts the BH in the Type I regime, see \citet{Duffellgap2015})
then the torque estimated by $T_0$ is $\sim3$ orders of magnitude weaker than the torque exerted
on our more massive, gap-opening secondary. 
However, it is currently unclear how the
torque changes with migration rate in this migration regime. Additionally, while
IMRIs will characteristically have higher SNR, the current
capabilities of numerical relativity are computationally limited for
calculating waveforms for intermediate mass ratio systems (\citealt{Mandel2009}). 
Accurate waveforms will be crucial for extracting the system parameters and for detecting a
gas-induced phase drift.

An important question to consider is whether the impact of gas may be
degenerate with system parameters (such as chirp mass, inclination,
eccentricity, spin, etc.) or with other environmental effects, such as
dynamical friction from dark matter
(\citealt{BarausseCardosoPani2015}). In principle, gas torques can
also be degenerate with modifications to general relativity, although
this effect would then be present in all E/IMRIs, while gas would have
a variable effect from source to source. An exploration of the
parameter space is beyond the scope of the present paper, and we leave
it to future work.  However, we note that the frequency dependence of
the gas torque is generally different from that of the system
parameters. At least for high SNR measurements for a
source that chirps over a broad frequency band, and for which the
frequency-dependence can be measured, we expect that the gas effects
can be disentangled from system parameter variations (see discussion
in \citealt{Yunes2011}).

We focus primarily on migration torques, but accretion discs can
produce other interesting effects on BH inspirals that we neglect in
the present study, including mass, spin, and eccentricity evolution.
Efficient accretion can lead to a non-negligible increase in mass of
the secondary BH throughout the inspiral, which will affect the GW
frequency evolution (accretion onto the primary SMBH is negligible, as
the increase in mass throughout the LISA lifetime is under the
accuracy limit).  The possibility exists for accretion to drive up the
spins of the BHs (e.g. \citealt{Teys2017}), an effect which we neglect
here, and it may also act to align the spins of both BHs, particularly
if the BHs form in the accretion disc \citep{Bogdanovic2007}.  While
we do not model the spin of the BHs in this paper, we hypothesise that
significant spin-up of the secondary is unlikely (at least for the
prograde case we simulated here), because the relative angular
momentum of the gas near the BH is low.  This is shown in the velocity
map in Fig.~\ref{fig:velocity}, where the gas around the BH exhibits
properties more akin to an atmosphere than to a mini-disc.

Gas also provides the possibility for EM counterparts to the GW detection. 
If an associated EM signature is detected, it would confirm the presence 
of gas around the source. In this case, even a non-detection of a GW phase deviation
for an EM-identified IMRI would teach us about the environment by putting a limit
on the density and/or viscosity of the disc.

Simulations of intermediate-mass planets show that a disc may drive
periodic eccentricity oscillations in the planet's orbit that are low
but not negligible, with the eccentricity ranging from 0.01 to 0.1
(\citealt{Ragusa2018}, see also \citealt{Papa2001, Bitsch2013})
although this depends on both the disc mass and the perturber mass
(\citealt{Dunhill2013}). We expect that gas-embedded E/IMRIs will be
close to circular compared to events originating from stellar remnants
in galactic nuclei, which will have significantly higher eccentricities.  Thus
spin-aligned and near-circular events will be indicative of a gas
disc.

In future work, we intend to relax several assumptions made in this
study, as well as explore a range of values for the key parameters.
In this paper we assumed a locally isothermal equation of state. This
assumption should be relaxed by incorporating a more sophisticated
treatment of the thermodynamics, allowing the gas to heat due to local
shocks and viscous dissipation, and to cool through its surface
\citep[e.g.][]{Farris+15}.  It is also known that the
$\alpha$-prescription is a poor approximation in parts of the disc
where the flow is not laminar, an issue that can be addressed using
magneto-hydrodynamics simulations~\citep[e.g.][]{Shi+2012}.  As
already mentioned above, a near-Eddington circumbinary accretion disc
around a BH in the LISA band is likely to be supported by radiation
pressure, which will need to be included in future simulations.
Finally, our simulations in this study are in 2D. We expect that the
3D vertical structure will modify the structure of the gap and the
accretion streams, as well as the gas distribution near the secondary
BH, and will inevitably have a strong affect on the gas torques.

We also expect that the migration torques depend on disc parameters
(such as $\alpha$ and $\mach$), the density and temperature profiles
of the disc, the BH mass ratio, any orbital eccentricity, and also the
GW-driven orbital decay rate.  Indeed, even for non-migrating planets,
\citet{Duffell2015} find that disc torques are sensitive to
combinations of these parameters, particularly in the intermediate
mass-ratio regime.

Despite its limitations, our present study suggests that the impact of
circumbinary gas may be measurable in the LISA waveform of an E/IMRI
event and warrants further investigation. We intend to address the
outstanding issues in future work.

\section{Conclusions}
\label{sec:summary}

In this paper, we study the gas torques exerted on a gravitational
wave driven inspiral with high resolution 2D hydrodynamic simulations
of a $10^{-3}$ mass ratio binary in an isothermal viscous
disc. Motivated by the prospect of LISA detecting the late stages of
intermediate mass ratio inspirals, we apply the results of our
simulations to estimate the detectability of gas torques on a $10^6
{\rm M_{\odot}}+10^3 {\rm M_{\odot}}$ binary merger occurring at a
redshift $z=1$.

We find that the net disc torque differs from previous semi-analytic
estimates, which were based on the viscous torque for a non-migrating
secondary. While these previous torque estimates were negative, we
here find that the total torque is positive, resulting in a slow-down
of the inspiral, and its strength is only fraction ($1\% - 5\%$) of
the viscous torque. While it is $4-5$ orders of magnitude weaker
than the torque due to GW emission, it can still produce a detectable
phase drift in the GW waveform. 
For our fiducial estimate of the accretion rate,
the accretion torque is at least $10-11$ orders of magnitude weaker
than that due to GWs and does not contribute significantly to the gas imprint in 
the waveform. 

An analysis of the origin of the torques shows that gas very close to the
secondary BH (inside its Hill sphere) exhibits a front-to-back
asymmetry with respective to the direction of the secondary's motion,
and leads to the positive (outward) component of the torque, whereas
gas elsewhere in the disc exerts a weaker negative torque.  More
sophisticated simulations that resolve the 3-dimensional gas
morphology  and velocity of gas near the secondary BH could provide insight into
whether and how this asymmetry occurs.

For the IMRI we consider here, the deviation in the GW waveform is
detectable (with a signal to noise ratio $> 10$) if the system is
embedded in a disc with a surface density $\Sigma_0 \gtrsim 10^3 \,
\rm g \, cm^{-2}$. This density may be exceeded in cold, thin,
near-Eddington discs expected in active galactic nuclei. Gas-induced
deviations are strongest during the earlier stages of the inspiral,
but they are more detectable for binaries at higher frequencies, where
LISA's sensitivity is stronger and the binary is chirping
significantly.  We expect the gas disc-induced phase drift in the GW
waveform to be sensitive to disc properties, which implies that the
detection of a gas-embedded inspiral will provide the opportunity for
LISA to probe the physics of AGN discs and migration torques.

The authors thank the anonymous referee for insightful comments and suggestions that improved this paper. 
AMD acknowledges support by the National Science Foundation (NSF)
Graduate Research Fellowship under Grant DGE 1644869.  The authors
acknowledge financial support from NSF grants DGE 1715661 and AST-1715356, NASA grants
NNX17AL82G, 16-SWIFT16-0015, and Einstein
Postdoctoral Fellowship award number PF6-170151 (DJD).

\bibliographystyle{mnras}

\bibliography{paper}

\end{document}